\documentclass[useAMS,usenatbib,usegraphicx]{mn2e}

\usepackage{times}

\newcommand{\singlecolsize}{0.45}
\newcommand{\middlecolsize}{0.80}
\newcommand{\doublecolsize}{0.90}

\newcommand{\ccfom}{{\rm cc_{fom}}}
\newcommand{\ccfomi}{\mathrm{cc}_{\mathrm{fom},i}}
\newcommand{\ccfomj}{\mathrm{cc}_{\mathrm{fom},j}}
\newcommand{\rmsmad}{\frac{\rm rms}{\rm mad}}
\newcommand{\sqdeg}{{\rm deg}^2}

   \newcommand{\aap}{A\&A}
\newcommand{\aj}{AJ}         \newcommand{\apj}{ApJ}
      \newcommand{\apjs}{ApJS}
\newcommand{\mnras}{MNRAS}   
     \newcommand{\pasp}{PASP}
\newcommand{\procspie}{Proc.\ SPIE}  

\newcommand{\new}[1]{{#1}} 

\title[GAMA: \textsc{autoz} redshift measurements]
{Galaxy And Mass Assembly (GAMA): \textsc{autoz} spectral redshift measurements,
  confidence and errors.}

\author[I.~K.~Baldry et al.]
{{\parbox{\textwidth}{\raggedright 
I.~K.~Baldry,$^1$
M.~Alpaslan,$^{2,3}$
A.~E.~Bauer,$^4$
J.~Bland-Hawthorn,$^5$
S.~Brough,$^4$
M.~E.~Cluver,$^6$
S. M. Croom,$^5$
L.~J.~M. Davies,$^2$
S.~P.~Driver,$^{2,3}$
M.~L.~P.~Gunawardhana,$^7$
B.~W.~Holwerda,$^8$
A.~M.~Hopkins,$^4$
L.~S.~Kelvin,$^9$
J.~Liske,$^{10}$
\'A.~R.~L\'opez-S\'anchez,$^{4,11}$
J.~Loveday,$^{12}$
P.~Norberg,$^7$
J.~Peacock,$^{13}$
A.~S.~G.~Robotham,$^2$
E.~N.~Taylor$^{14}$
}}\\
\vspace{0.4cm}\\
{\parbox{\textwidth}{\raggedright 
$^1$Astrophysics Research Institute, Liverpool John Moores University,
IC2, Liverpool Science Park, 146 Brownlow Hill, Liverpool L3 5RF 
\\
$^2$International Centre for Radio Astronomy Research, 
University of Western Australia, 35 Stirling Highway, Crawley, WA 6009, Australia 
\\
$^3$School of Physics and Astronomy, University of St Andrews, 
North Haugh, St Andrews KY16 9SS
\\
$^4$Australian Astronomical Observatory, PO Box 915, North Ryde, NSW 1670, Australia
\\
$^5$Sydney Institute for Astronomy, School of Physics A28, University of Sydney, NSW 2006, Australia
\\
$^6$Department of Astronomy, University of Cape Town, Private Bag X3, Rondebosch, 7701, South Africa 
\\
$^7$ICC, Department of Physics, Durham University, South Road, Durham DH1 3LE 
\\
$^8$University of Leiden, Sterrenwacht Leiden, Niels Bohrweg 2, 
NL-2333 CA Leiden, Netherlands 
\\
$^9$Institut f\"{u}r Astro- und Teilchenphysik, Universit\"{a}t Innsbruck, Technikerstra{\ss}e 25, 
6020 Innsbruck, Austria 
\\
$^{10}$European Southern Observatory, Karl-Schwarzschild-Str.~2, 85748 Garching, Germany
\\
$^{11}$Department of Physics and Astronomy, Macquarie University, NSW 2109, Australia 
\\
$^{12}$Astronomy Centre, University of Sussex, Falmer, Brighton BN1 9QH 
\\
$^{13}$Institute for Astronomy, University of Edinburgh, Royal Observatory, Blackford Hill, Edinburgh EH9 3HJ 
\\
$^{14}$School of Physics, University of Melbourne, Parkville, VIC 3010, Australia
\\
}}}

\begin{document}

\date{Paper submitted to MNRAS on 2014 February 26th; revised following
  referee comments, 2014 April 8th.}

\pagerange{\pageref{firstpage}--\pageref{lastpage}} \pubyear{2014}

\maketitle

\label{firstpage}

\begin{abstract}
  The Galaxy And Mass Assembly (GAMA) survey has obtained spectra of over
  230\,000 targets using the Anglo-Australian Telescope. To homogenise the
  redshift measurements and improve the reliability, a fully automatic
  redshift code
  was developed (\textsc{autoz}).  The measurements were made using a
  cross-correlation method for both absorption-line and emission-line
  spectra. Large deviations in the high-pass filtered spectra are partially
  clipped in order to be robust against uncorrected artefacts and to reduce
  the weight given to single-line matches. A single figure of merit (FOM) was
  developed that puts all template matches onto a similar confidence
  scale. The redshift confidence as a function of the FOM was fitted with a
  tanh function using a maximum likelihood method applied to repeat
  observations of targets. The method could be adapted to provide robust
  automatic redshifts for other large galaxy redshift surveys.
  For the GAMA survey, there was a substantial improvement in the
  reliability of assigned redshifts and in the lowering of redshift
  uncertainties with a median velocity uncertainty of 33\,km/s.
\end{abstract}


\section{Introduction}
\label{sec:intro}

Spectroscopic redshift measurements of large galaxy samples form the backbone
of many extragalactic and cosmological analyses.  They are key for testing
cosmological models, e.g.\ using redshift space distortions \citep{Kaiser87},
and for providing distances for galaxy population studies when the cosmology
is assumed.
Redshifts from spectroscopy (spec-$z$) generally have significantly fewer
outliers, compared to the true redshift, than from photometric estimates
(photo-$z$; \citealt{dahlen13}).  In addition, spec-$z$ measurements are
essential for accurate low-redshift luminosity estimates ($0.002 \la z \la
0.2$), where the photo-$z$ fractional error is too large, and for dynamical
measurements within groups of galaxies 
(\citealt*{BFG90}, \citealt{robotham11}).

Redshift surveys of large numbers of galaxies have been undertaken in recent
years using multi-object spectrographs such as the Two Degree Field (2dF,
\citealt{lewis02df}), Sloan Digital Sky Survey (SDSS, \citealt{smee13}),
Visible Multi-Object Spectrograph (VIMOS, \citealt{lefevre03}), and Deep
Imaging Multi-Object Spectrograph (DEIMOS, \citealt{faber03}).  For uniformity
of a survey product over a large sample, redshift measurement codes have been
developed that are either fully automatic
\citep{subbarao02,garilli10,bolton12} or partially automatic with some user
interaction \citep{colless01,newman13}.

The main techniques for spectroscopic redshift measurements are: the
identification and fitting of spectral features \citep{MW95};
cross-correlation of observed spectra with template spectra
\citep{TD79,kurtz92}; and $\chi^2$ fitting using linear combinations of
eigenspectra \citep*{GOD98}.  A widely used code is \textsc{rvsao} that allows
for cross-correlation with absorption-line and emission-line templates
separately \citep{KM98}.  The 2dF Galaxy Redshift Survey (2dFGRS) and SDSS
have used a dual method with fitting of emission line features and
cross-correlation with templates after clipping the identified emission lines
from the observed spectra \citep{colless01,stoughton02}.  The large VIMOS
surveys have used the \textsc{ez} software \citep{garilli10}, which provides a
number of options including emission-line finding, cross-correlation and
$\chi^2$ fitting.  From SDSS Data Release 8 (DR8) onwards \citep{sdssDR8},
including the Baryon Oscillation Spectroscopic Survey (BOSS) targets, the
measurements have used $\chi^2$ fitting at trial redshifts with sets of
eigenspectra for galaxies and quasars \citep{bolton12}.



The Galaxy And Mass Assembly (GAMA) survey is based around a redshift survey
that was designed, in large part, for finding and characterising groups of
galaxies \citep{driver09,robotham11}.  The survey has obtained over 200\,000
redshifts using spectra from the AAOmega spectrograph of the Anglo-Australian
Telescope (AAT) fed by the 2dF fibre positioner.  AAOmega is a bench-mounted
spectrograph with light coming from a 392-fibre slit, split into two beams,
each dispersed with a volume-phase holographic grating and focused onto CCDs
using a Schmidt camera. See \citet{sharp06} for details.

Up until 2013, all the AAOmega redshifts had been obtained using \textsc{runz}
\citep*{SCS04}, which is an update to the code used by the 2dFGRS
\citep{colless01}.  The user assigns a redshift quality for each spectrum from
1--4, which can later be changed or normalised during a quality control
process \citep{driver11}.  By comparing the redshifts assigned to repeated
AAOmega observations of the same target, the typical redshift uncertainty was
estimated to be $\sim100\,$km/s. In addition, the blunder rate was $\sim5$
per cent even when the redshifts were assigned a reliable redshift quality of
4. In order to improve the redshift reliability and uncertainties, and thus
the group catalogue measurements, a fully automatic code was developed called
\textsc{autoz}.

Here we describe the \textsc{autoz} algorithm, which uses a cross-correlation
method that works equally well with absorption or emission line templates, and
that is robust to additive/subtractive residuals and other uncertainties in
the reduction pipeline that outputs the spectra.  This has substantially
improved the GAMA redshift reliability and velocity errors (Liske et al.\
in preparation). A description of the GAMA data is given in 
\S~\ref{sec:data}. The method for finding the best redshift estimate
is outlined in \S~\ref{sec:method}, the quantitative assessment
of the confidence is described in \S~\ref{sec:confidence},
and the redshift uncertainty estimate is described in
\S~\ref{sec:uncertainty}. A summary is given in \S~\ref{sec:summary}. 

\section{Data}
\label{sec:data}

The GAMA survey is based around a highly complete galaxy redshift survey and
multi-wavelength database \citep{driver11}.  Since the initial spectroscopic
target selection over 144$\,\sqdeg$ described in \citet{baldry10}, the survey
has been expanded to 280$\,\sqdeg$ with a main survey limit of $r<19.8$ in all
five regions: equatorial fields G09, G12 and G15 and Southern fields G02 and
G23.  The spectra were obtained from the Anglo-Australian Telescope with the
Two-degree Field (2dF) robotic positioner fibre feed to the AAOmega
spectrograph. In total, 286\,705 spectra \new{of 237\,822 unique targets were taken over six
years, in all weather conditions}. The spectra were reduced using
\textsc{2dfdr} \citep*{CSH04}. 
The GAMA setup and data processing details are described in \citet{hopkins13}.

In order to obtain high completeness, both in targeting and redshift success,
the same region of the sky was observed multiple times with different AAOmega
configurations \citep{robotham10}.  The spectra were reduced using
\textsc{2dfdr} and redshifts were measured using \textsc{runz} typically
within 24 hours of the observations. In between observing seasons, the spectra
were usually re-reduced using the most recent update to the \textsc{2dfdr}
pipeline processing. In addition, a significant fraction of spectra were 
looked at by multiple users in order to quantify the 
reliability of the \textsc{runz} redshift assignments (Liske et al.\ in 
preparation). The tiling catalogue was updated during an observing
season, or in between observing seasons, using the latest assigned redshift
qualities. If a target was not assigned a sufficient redshift quality and had
not been observed twice by AAOmega, then it remained at high priority in the
tiling catalogue. In most cases, a target received a higher-quality redshift
on its second attempt.

\begin{figure}
\centerline{
\includegraphics[width=\singlecolsize\textwidth]{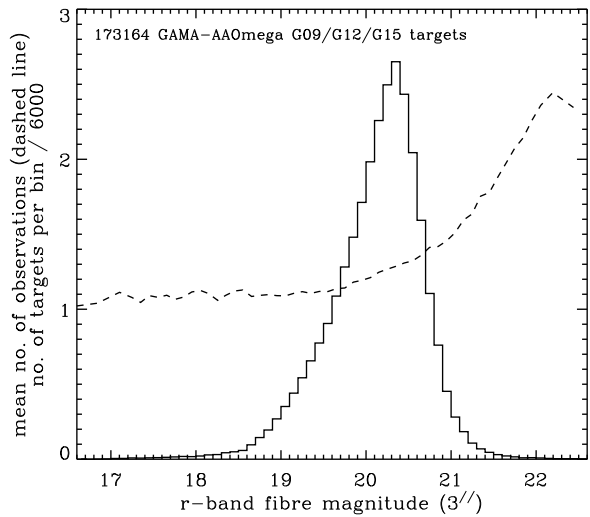} 
}
\caption{The distribution in the fibre magnitude of targets observed
  as part of the GAMA-AAOmega campaign (solid line), 
  and the mean number of observations
  as a function of fibre magnitude (dashed line), for the equatorial fields. 
  The fibre magnitudes were obtained from the SDSS catalogue.}
\label{fig:fibre-histo}
\end{figure}

Fig.~\ref{fig:fibre-histo} shows the distribution of fibre magnitudes of the
GAMA targets that were observed by AAOmega (most targets with high quality
redshifts from SDSS or 2dFGRS were not observed), and the mean number of
AAOmega observations as a function of fibre magnitude. At bright fibre
magnitudes, only a few per cent of targets were observed twice because they
were first observed in poor conditions or for quality control purposes. At
fainter fibre magnitudes, the number of targets observed more than once is
higher because, even in average or good conditions, the \textsc{runz} user was
assigning a low quality redshift. Thus the strategy has worked, as expected,
to obtain more observations for the fainter targets. As a result of the
strategy, we have repeat observations of 40\,319 targets.  These repeats have
been used to fine tune a new fully automatic redshift code called
\textsc{autoz}, and can be used for coadding to increase the spectral
signal-to-noise ratio. This has significantly improved the redshift
measurements with respect to the user assigned redshifts from \textsc{runz}
(Liske et al.\ in preparation).

\section{Redshift measurements}
\label{sec:method}

The code \textsc{autoz} uses the cross-correlation method for obtaining
redshifts for spectra with or without strong emission lines.
It is described as a {\it fait accompli}, but there have been
several iterations, each time checking to see if changes to the code
resulted in an improvement. The tuning of parameters 
was optimised for performance with the AAOmega spectra. 
After each iteration, for example, redshift
confidence estimates were calibrated 
from the repeat observations, the total number of high
confidence redshifts was determined, and selected spectra 
and cross-correlation functions were inspected. 

\subsection{Spectral templates}

The SDSS has set a high standard for automatic redshift determination.  For
\textsc{autoz}, we used their templates for spectral cross-correlation.  These
are a high S/N set of coadded spectra given in a similar format to the 
SDSS spectra for scientific targets.  
Table~\ref{tab:templates} lists the templates used for this
paper. Twenty stellar spectra were used. \new{Early versions of \textsc{autoz}
used the six SDSS DR2 galaxy templates \citep{subbarao02}, while
later versions used eight galaxy
templates that were created from the SDSS-BOSS galaxy eigenspectra
\citep{bolton12}.}  Six of these templates were chosen to closely match the DR2
templates where there was common wavelength coverage, with an additional two
selected to represent a post-starburst spectrum \citep{wild07} and a typical
SDSS-BOSS spectrum.

\begin{table*}
\caption{Spectral templates used for cross correlation.
  The O-star and L1-star stellar templates were not included because they have
  negligible chance of genuine matches with our GAMA sample. For the 
  \textsc{autoz} redshifts used by the GAMA team, 
  the initial database versions (SpecCat v20, v21) used 20 stellar and 6 galaxy templates (23--28), 
  while later versions used 20 stellar and 8 galaxy templates (40--47).
  In either case, the galaxy templates cover a range of emission-line strengths relative to
  the absorption lines.}
\label{tab:templates}
\begin{tabular}{llllll} \hline 
template numbers & file & spectral types& rest-frame $\lambda/$\AA & search
$z$ range & noise-estimate $z$ range \\ \hline
02--10 & spDR2-...            & B to K stars    & 3800--9150 & $-$0.002 to 0.002 & $-$0.1 to 0.5 \\
11--14, 17, 19, 22 & spDR2-... & late-type stars & 3800--9150 & $-$0.002 to 0.002 & $-$0.2 to 0.4 \\
16, 18, 20, 21 & spDR2-...    & other stars     & 3800--9150 & $-$0.002 to 0.002 & $-$0.1 to 0.5 \\
23--27 & spDR2-...            & galaxies        & 3500--9000 & $-$0.005 to 0.8   & $-$0.1 to 0.8 \\
28 & spDR2-028            & luminous red galaxy & 3000--6800 & $-$0.005 to 0.8   & $-$0.1 to 0.8 \\ 
40--47 & spEigenGal-55740     & galaxies        & 2500--9000 & $-$0.005 to 0.9   & $-$0.1 to 0.9 \\ 
\hline
\end{tabular}
\end{table*}

The spectra were available rebinned onto a vacuum wavelength scale
separated by 0.0001 in $\log_{10}\lambda$. This corresponds to a pixel
size of 0.92\,\AA\ at 4000\,\AA\ and 1.84\,\AA\ at 8000\,\AA.

\begin{figure*}
\centerline{
\includegraphics[width=\doublecolsize\textwidth]{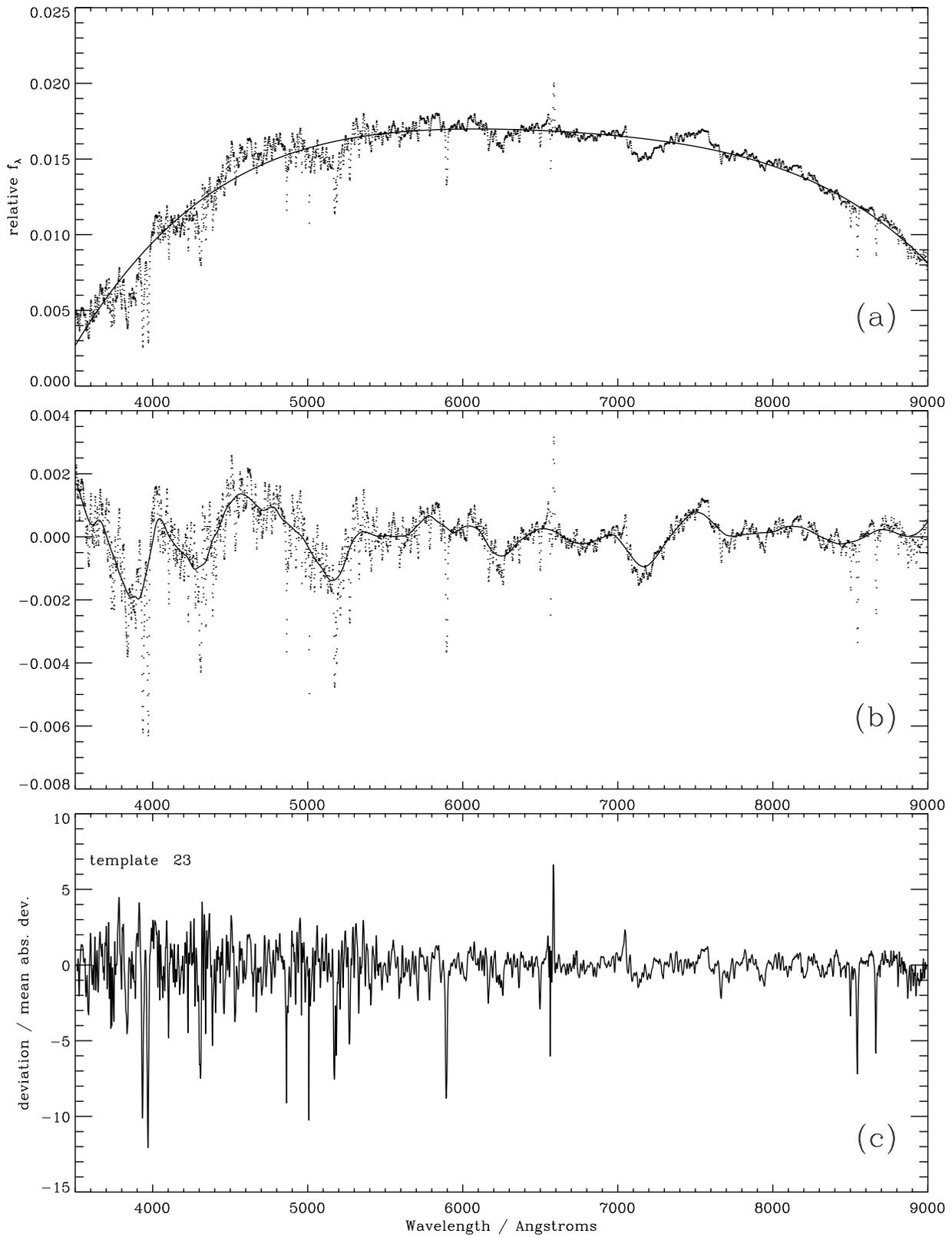} 
}
\caption{The high-pass filtering procedure for template 23. (a) The
  template spectrum is shown by the points, and the initial polynomial
  fit by the line. (b) The spectrum subtracted by the polynomial is
  shown with points, and the median filtered and smoothed version is
  shown by the line. (c) The final HPF template spectrum.}
\label{fig:hp-procedure}
\end{figure*}

The spectral templates were high-pass filtered using a two step
robust procedure:
\begin{enumerate}
\item A 4th-order polynomial was fitted to the spectrum iteratively,
  with a maximum of 15 iterations.  After each iteration, points more
  than 3.5$\sigma$ away from the best-fit curve were rejected. 
  The final 4th-order polynomial was then subtracted from the spectrum. 
\item A median kernel filter of width 51 was applied to the result of
  the 1st step. On each end, the 25 edge points were given a median value
  of those points.  This median filtered spectrum was
  smoothed using a trapezium filter, by applying two boxcar smooths of width
  121 and 21. This low-pass spectrum was then subtracted from the
  result of the 1st step to obtain the high-pass filtered (HPF) 
spectrum.
\end{enumerate}
The aim of the first step was to remove the large-scale modes from the
spectrum so that the second step involving median filtering was not
compromised by a steeply rising spectrum, for example. 
The HPF spectra also had their edge points set smoothly to
zero using a cosine bell taper (apodization, \citealt{KM98}).  The chosen or
maximum wavelength coverage of each template is given in
Table~\ref{tab:templates}.
A demonstration of the
method is shown in Fig.~\ref{fig:hp-procedure} for the old-stellar galaxy
template number 23. 

\begin{figure*}
\centerline{
\includegraphics[width=\doublecolsize\textwidth]{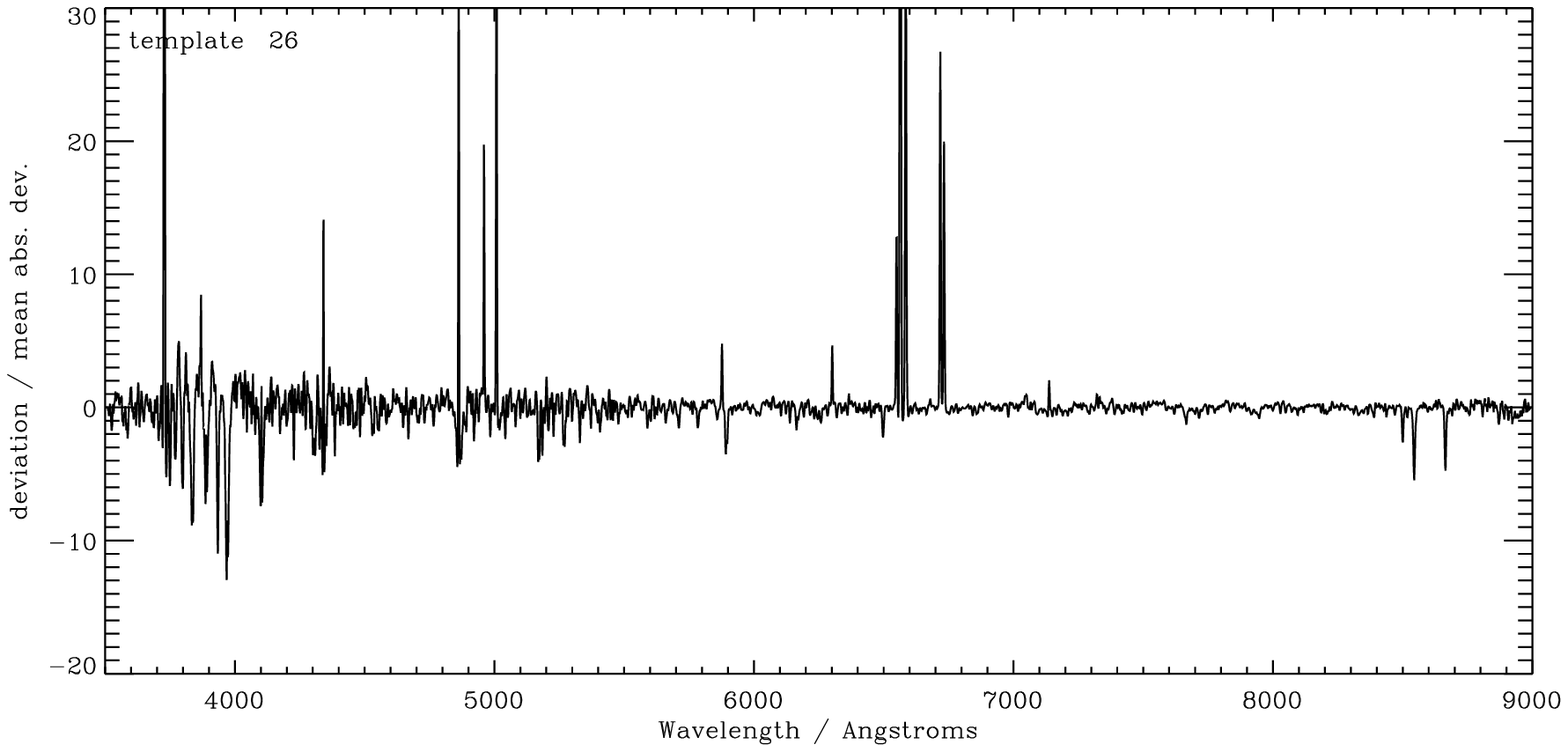} 
}
\caption{The high-pass filtered spectrum for template 26. The emission
lines are clipped to 30 times the mean absolute deviation.}
\label{fig:template26}
\end{figure*}

The final HPF spectra for the templates were clipped to lie between $-30$ and
$+30$ times the mean absolute deviation, determined iteratively until
convergence within a small tolerance.  An example of this is shown in
Fig.~\ref{fig:template26}. Here five emission lines are clipped. This is to
avoid giving too much weight to a single strong line, which could give rise to
cross-correlation spikes in bad data. In effect, a cross-correlation of this
template with data gives a high peak when two or more lines line up with
matched wavelength spacing.  In other words, the correct wavelength spacing
gives rise to higher confidence in the redshift with less weight given to the
relative strengths of any lines.

We note that in SDSS DR8 onwards, the method for determining redshifts
used $\chi^2$ fitting with combinations of four eigenspectra for
galaxies \citep{bolton12}. While we cannot follow their method for
fitting because of the less reliable spectrophotometry of the AAOmega
spectra, combinations of these eigenspectra were used to extend the
wavelength range of the galaxy templates down to shorter
wavelengths. These are templates 40--47 in the \textsc{autoz} code, in
order of increasing strength of emission lines.\footnote{We have
  chosen not to use quasar templates at this stage because of the
  significantly different scale and frequency of the features; they
  probably account for less than one per cent of the GAMA main sample.
  The confidence estimate for quasar redshifts is more difficult to
  make automatically when there is only one or two broad emission
  lines across the observed wavelength range. \new{In addition, 
  AAOmega spectra sometimes show an unfortunate broad artefact
  at the join between the red and blue arms, which may be confused with
  a broad line}.}

\subsection{AAOmega spectra}
\label{sec:aaomega-spectra}

The AAOmega pipeline produces a spectrum and error spectrum for each target,
with the red and blue beams combined to a single linear scale with a pixel
width of 1.036\,\AA. These were then approximately flux calibrated to relative
$f_\lambda$ units using an average flux correction determined for the
survey. The reason for doing this is that the SDSS spectral templates are
calibrated in $f_\lambda$ units; and even though the spectra are high-pass
filtered, the weighting is affected by the calibration.\footnote{We do not use
  the flux calibration applied to each AAOmega configuration separately
  \citep{hopkins13}.  This fails in a few per cent of cases and may introduce
  incorrect flux variations across the spectra in other cases.  Instead a
  robust, quadratic function, average flux calibration was determined and
  applied to all the unflux-calibrated spectra. In any case, flux calibration
  is not critical for cross-correlation and only affects the relative
  weighting between different parts of a spectrum.}

\begin{figure*}
\centerline{
\includegraphics[width=\doublecolsize\textwidth]{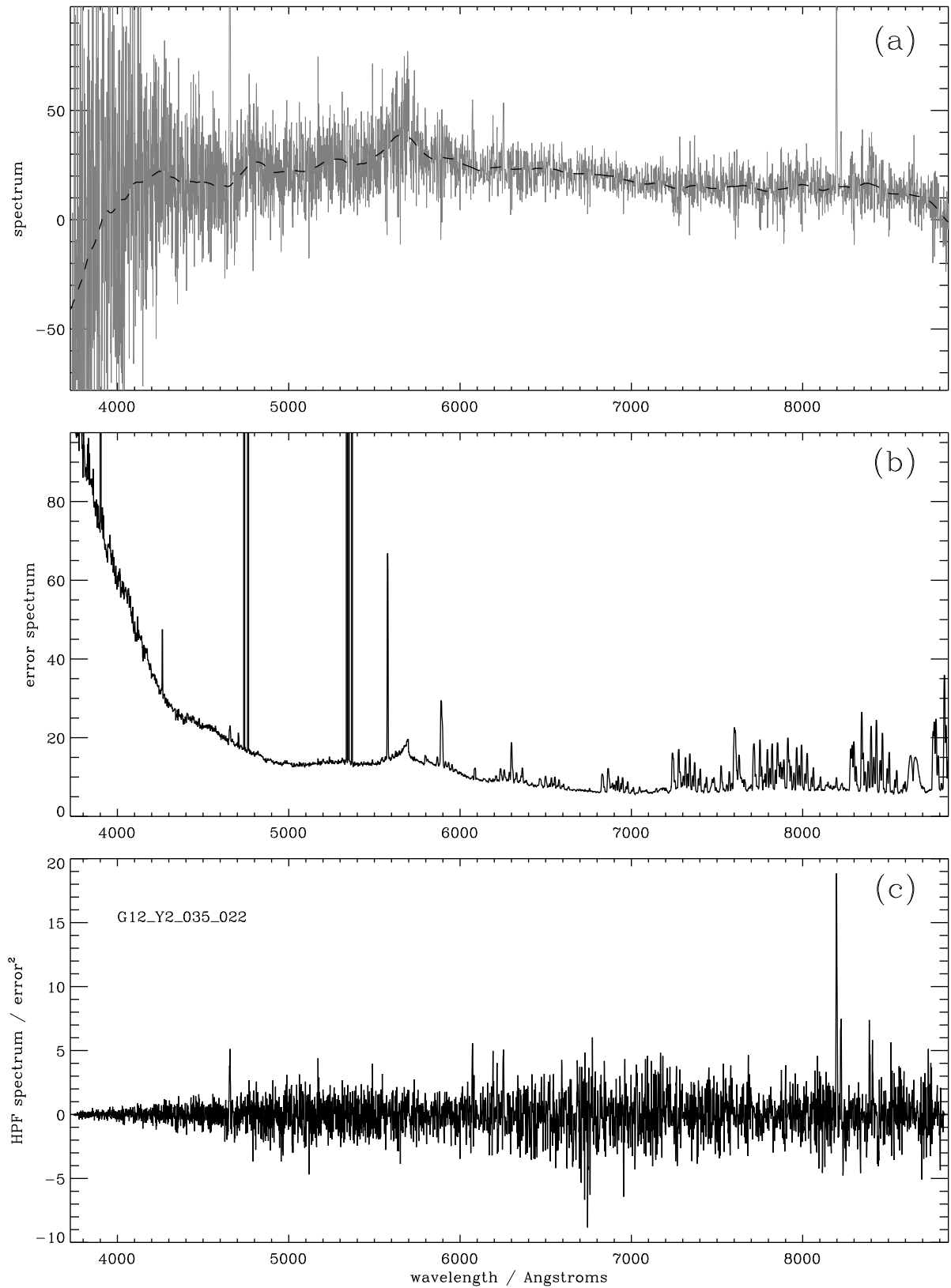} 
}
\caption{The high-pass filtering procedure for the AAOmega
  spectra. (a) The spectrum is shown in grey with the low-pass
  spectrum, to be subtracted from the spectrum, shown by the dashed
  line.  (b) The uncertainty in the spectrum. Known bad pixels are 
  represented by off-the-scale values giving rise to the spikes. (c) The final
  HPF spectrum to be used to produce cross-correlation functions.}
\label{fig:example-hpf}
\end{figure*}

The AAOmega spectra for the GAMA survey were high-pass filtered in the same
way as the spectral templates.  The cosine bell taper was set between 3786 and
3736\,\AA\ at the low-wavelength end and between 8790 and 8840\,\AA\ at the
high-wavelength end.  Each error spectrum was broadened using a maximum filter
kernel of width 3.  This broadening allows for the uncertainty in the alignment
of sky subtraction. In other words, this is to account for underestimation by
\textsc{2dfdr} of the error spectrum near sky lines.
The HPF spectrum was then divided by the square of the error spectrum.  
This weighting was justified by \citet{SCS04} as 
appropriate for effectively minimising $\chi^2$ when finding the peak. 
Fig.~\ref{fig:example-hpf} shows the filtering procedure applied to an AAOmega
spectrum.

The high-pass filtering is aggressive for both the templates and
AAOmega spectra. This is to mitigate against artefacts on the scale of
around 200\,\AA\ and longer, for example, `fringing' caused by a
separation between the prism and optical fibre at the 2dF plate, and
an imperfect join between the spectra from the red and blue arms of
the spectrograph (see \citealt{hopkins13} for details). 
It is more straightforward to define a reliable automatic
confidence estimate using HPF spectra where these scales have been
removed.

The HPF spectra were then clipped to lie between $-25$ and 25 times the mean
absolute deviation.\footnote{The clipping limits were initially $\pm30$,
which were the same as those used for the templates. They were changed to $\pm25$
to alleviate some redshift disagreements between matched spectra at high
figure-of-merit values (\S~\ref{sec:confidence}).} 
This was chosen so that, in general, only high S/N spectra
had genuine lines clipped.  In these cases, the clipping does not compromise
the cross-correlation peak position significantly because there is more than
sufficient signal in less extreme deviations in any case.  The clipping of the
templates and AAOmega HPF spectra means that no single emission line, or
apparent emission line, can result in a high confidence redshift. 
\new{This approach is therefore more robust} to bad data: 
uncorrected hot pixels, cosmic rays, or
misaligned sky subtraction.  Bad pixels that had been accounted for were set
to zero in the HPF spectra.

\subsection{Cross-correlation functions}
\label{sec:cross-corr}

The template and target HPF spectra were linearly rebinned onto a
logarithmic vacuum wavelength\footnote{When converting the AAOmega
  spectra to a vacuum wavelength scale, the formula on the SDSS web
  pages was used: $\lambda_{\rm air} = \lambda_v / (1.0 + 2.735182
  \times 10^{-4}
  + 131.4182 / \lambda_v^2 + 2.76249 \times 10^8 / \lambda_v^4)$.  A lookup
  table was created for the conversion factor as a function of
  $\lambda_{\rm air}$, and the values determined for the AAOmega 
  wavelengths by interpolation. The conversion factor varies 
  between 1.000275 and 1.000285.} 
scale from 3.3 to 4.0 (with zero padding, \citealt{KM98})
with a pixel width of $2\times10^{-5}$, which corresponds to
$13.8{\rm\,km/s}$. For each target HPF spectrum, the cross-correlation
function was determined for all the templates using the usual
procedure involving fast Fourier transforms \citep{Simkin74}.  
\new{The cross-correlation values were associated with heliocentric redshifts given 
by
\begin{equation}
z_{{\rm ccf},i} = 
10^{(2 \times 10^{-5} \, \delta_{{\rm pix},i})} 
\left( 1+z_{\rm t} \right) 
\left( 1+\frac{v_{\rm sun,c}}{c} \right) - 1
\end{equation}
where $\delta_{{\rm pix},i}$ is the shift in pixels corresponding to position $i$; 
$z_{\rm t}$ is the redshift of the template spectrum, usually zero; 
and $v_{\rm sun,c}$ is component of the velocity of the Earth in the 
heliocentric frame toward the target (if the target spectrum 
was not put on a heliocentric wavelength scale). 
}

\begin{figure*}
\includegraphics[width=\middlecolsize\textwidth]{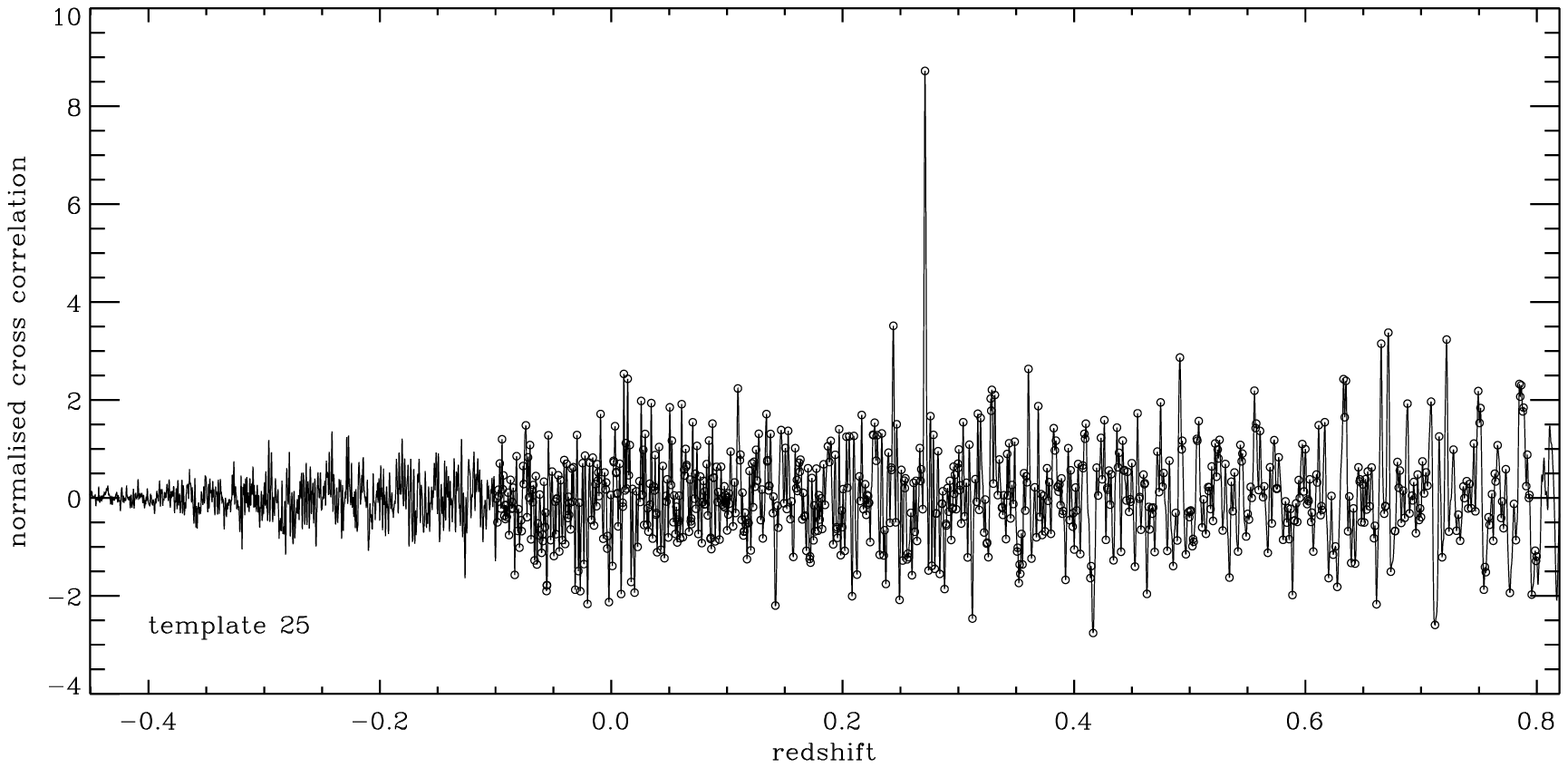} 
\includegraphics[width=\middlecolsize\textwidth]{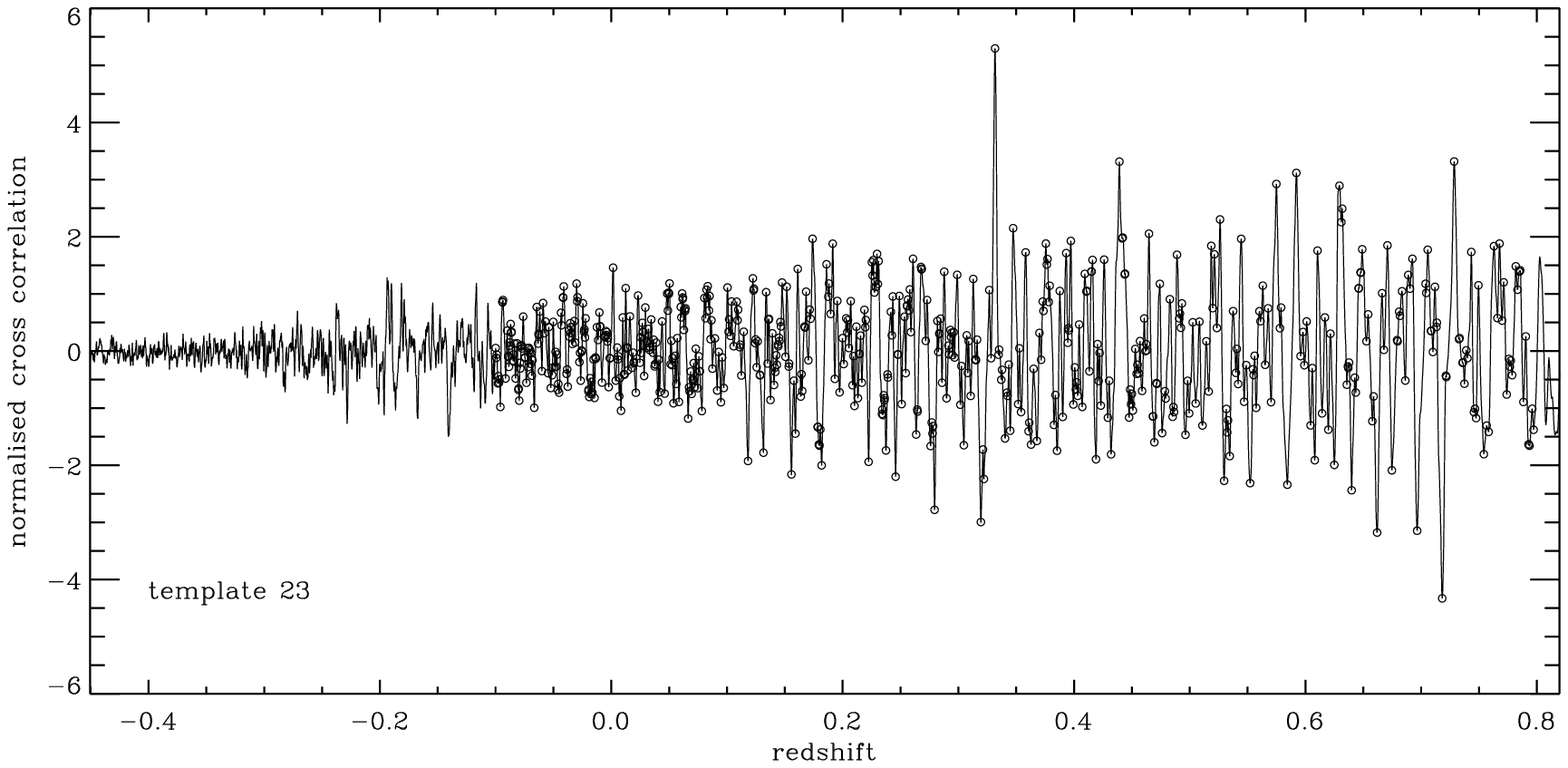} 
\includegraphics[width=\middlecolsize\textwidth]{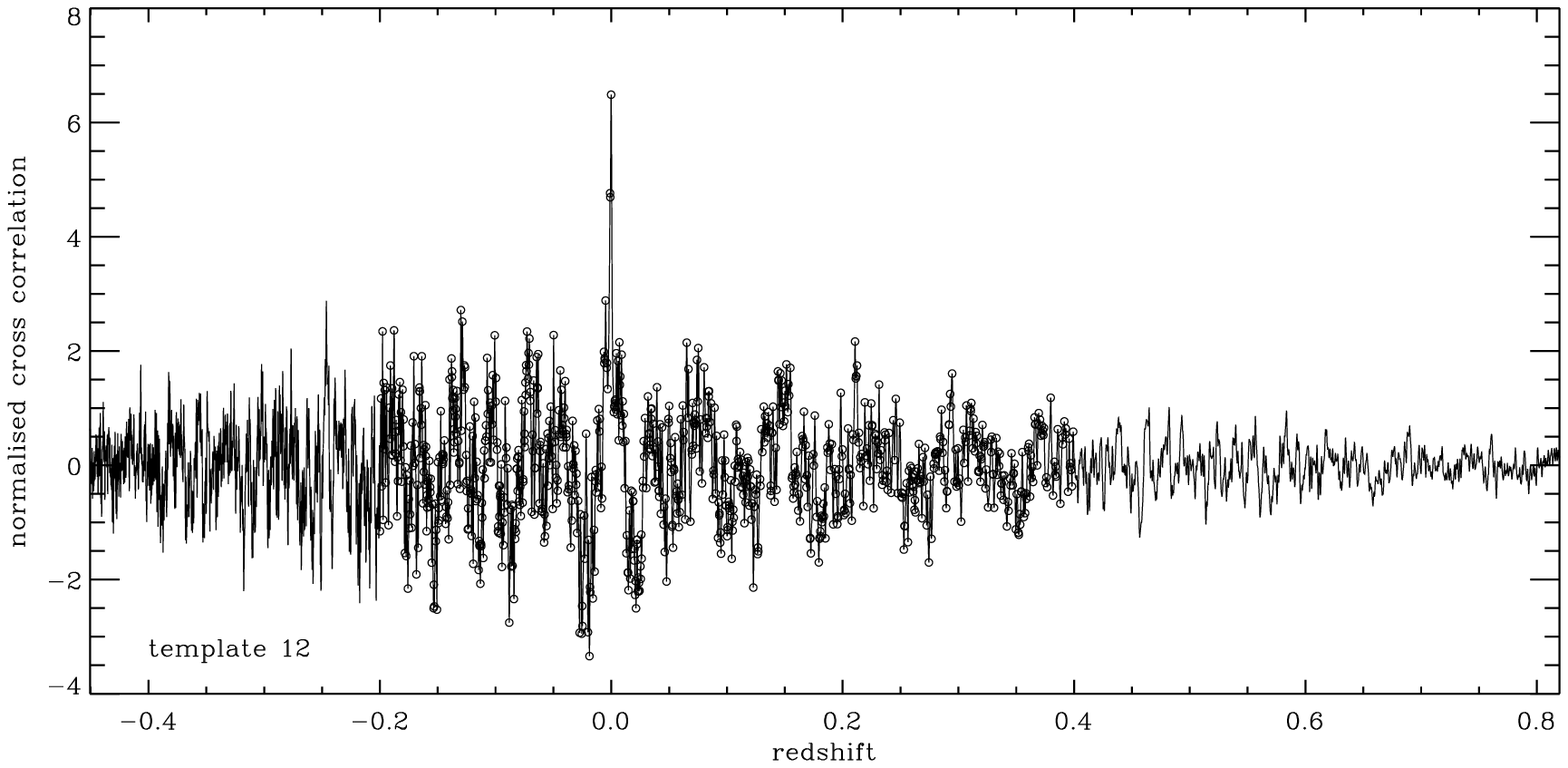} 
\caption{Examples of cross-correlation functions. The circles show the
turning points and range used for the normalisation.}
\label{fig:cross-correlation}
\end{figure*}

For each template, a search range for finding redshifts and a
noise-estimate redshift range \new{for the normalisation procedure} 
were defined.  The ranges for each template are
given in Table~\ref{tab:templates}.  The cross-correlation functions
were normalised by subtracting a truncated mean \new{computed} over the
noise-estimate range, which results in a small adjustment, and
dividing by the root mean square (RMS) of the turning
points \new{computed over the same range}.
Typically the cross-correlation functions had between 350 and
850 turning points over the noise-estimate
range. Fig.~\ref{fig:cross-correlation} shows three examples of
normalised cross-correlation functions.

For each \new{target} spectrum, 
all the normalised cross-correlation functions were
determined corresponding to each template being used
(26 or 28 templates; see Table~\ref{tab:templates}).  First, the highest
cross-correlation peak, within the search ranges, was taken to be the best
estimate of the redshift. The next three best redshifts were determined after
excluding peaks within 600\,km/s of the better redshift estimates, considering
all templates, at each stage.  The values of the peaks are called $r_x$
(following the nomenclature of \citealt{cannon06}),
$r_{x,2}$, $r_{x,3}$ and $r_{x,4}$ each with a corresponding redshift 
and template number.  In
order to avoid discretisation at the rebinned resolution, the redshifts were
fine tuned by fitting a quadratic to seven points centred on each peak.
\new{Note that the highest redshift allowed was set at 0.9, 
which is appropriate for the GAMA magnitude limit and the fact that 
the \textsc{autoz} code has not been adapted to search for quasar redshifts.}


\subsubsection{Note on the redshift range used for the normalisation}

\new{In order that the normalised cross-correlation functions,
$r_x$ values, can be
compared between different templates, it is important that the
noise-estimate ranges are set appropriately.  The cross-correlation
function computed using fast Fourier transforms gives values for
$z_{{\rm ccf},i}$ from $10^{-0.35} -1$ to $10^{0.35} - 1$ (about
$-0.55$ to 1.24) because of the rebinned logarithmic scale from 3.3 to
4.0.  Zero padding is necessary because of the wrap-around assumption
of this cross-correlation method. As a result, the amplitude of a
cross-correlation function can drop significantly when the overlap
between the rest-frame wavelength range of the template
(Table~\ref{tab:templates}) and the observed wavelength range of the
target decreases. Thus a more useful estimate of the noise is obtained
by computing the RMS over a reduced redshift range, here called the
noise-estimate range. The noise-estimate range for the galaxy
templates is chosen to encompass the search redshift range with
additional negative redshifts. The noise-estimate range for the
stellar templates covers a shorter range because of the reduced
rest-frame wavelength coverage. For late-type stars, the noise estimate
uses more negative redshifts because there are larger deviations at
the red-end of the template spectra.  As a final test of the
normalisation procedure, using repeated observations or otherwise, it
can be apparent if certain templates give rise to numerous false
redshift peaks (\S~\ref{sec:confidence}).}

\section{Redshift confidence estimation}
\label{sec:confidence}

In this section, we discuss the process for estimating the 
likelihood that the highest
normalised cross-correlation peak corresponds to the correct redshift.
In cases where the distribution of the cross-correlation function
values of the turning points is close enough to Gaussian, then the
value of $r_x$ can be used to give an estimate of the redshift
confidence (e.g.\ \citealt{cannon06}; see \citealt{Heavens93}
for a theoretical estimate). 
However, in most cases, there are a few high values of the 
cross-correlation functions because of aliasing between emission 
lines, for example. 
To test the reliability of the best redshift estimate, we also consider
the ratio between the highest peak and the subsequent peaks, here, given by
{\small
\begin{equation}
r_{x,\mathrm{ratio}} = \frac{r_x}
{\sqrt{ (r_{x,2})^2  + (r_{x,3})^2 + (r_{x,4})^2}} \mbox{~~.}
\label{eqn:ccsigma1to234}
\end{equation}
}

\begin{figure}
\centerline{
\includegraphics[width=\singlecolsize\textwidth]{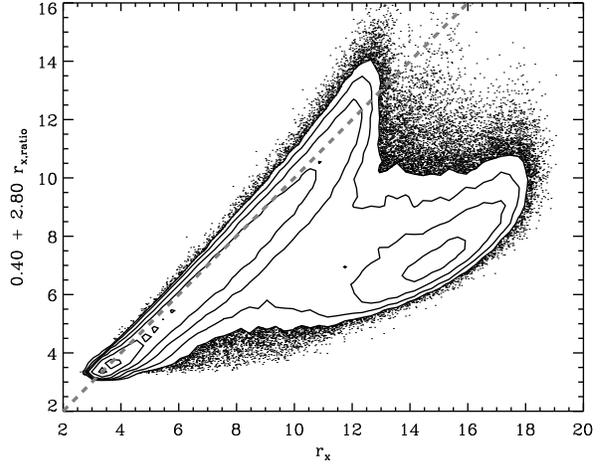} 
}
\caption{Bivariate distribution of $a_0 + a_1 \, r_{x,\mathrm{ratio}}$ versus
  $r_x$. The solid lines represent logarithmically spaced density
  contours with a factor of two between each level. The dashed line
  shows where the two values are equal.}
\label{fig:fom-calc}
\end{figure}

Fig.~\ref{fig:fom-calc} shows the distribution of an adjusted
$r_{x,\mathrm{ratio}}$ versus $r_x$ for 286\,705 AAOmega spectra. 
Cross-correlations with the templates that have weak or no emission
lines tend to follow the diagonal ridge line while spectra that have higher
peaks using the emission lines dominate the hill to the right. 
Cross-correlations with noise populate the bottom left corner. 
Of the two variables, $r_{x,\mathrm{ratio}}$ gives a better
figure of merit (FOM) for relating to redshift confidence. A slight
improvement can be made by defining the following:
\begin{equation}
  {\rm cc_{fom,prelim}} = \min( {r_x}, \,
  a_0 + a_1 \, r_{x,\mathrm{ratio}} ) \mbox{~~.}
\label{eqn:fom-prelim}
\end{equation}
with $a_0=0.4$ and $a_1=2.8$, and where the $\min()$
function returns the lower of the two variables.\footnote{The slope of 
the line $a_1$ was determined by fitting to the ridge line. 
The value of $a_0$ was adjusted by trial and error to 
give the largest number of high confidence redshifts from the AAOmega
spectra, after calibrating the confidence each time.
In the absence of sufficient repeat spectra, fitting to the ridge line would 
provide an adequate estimate for an improved FOM.} The line 
where these variables are the same is shown in Fig.~\ref{fig:fom-calc}. 

Three further adjustments, \new{penalty functions}, to the FOM were made. 
The first is because in cases
of poor sky subtraction or other reduction problems, ${\rm cc_{fom,prelim}}$
can still be above 5 even when there is no signal from the target source. For
each HPF spectrum, the ratio of the RMS to mean absolute deviation (MAD) was
determined. This value (rms/mad) is high when there is good signal from the
target or when there is non-Poisson noise, i.e., reduction problems. Thus, the
FOM can be reduced when ${\rm rms/mad} > 1.8$ (the median value is 1.44) without
losing genuine redshifts. The adjustment is given by
\begin{equation}
{\rm adjust_1} = 
\begin{array}{l} - 2.1 \\ - 1.5 (\rmsmad - 1.8) \\ ~~~~0 \end{array}
\mbox{~~~for~~~}
\begin{array}{l} \rmsmad > 3.2 \\ 1.8 < \rmsmad < 3.2 \\ \rmsmad < 1.8 
\end{array}
\mbox{~~~.}
\end{equation}

The second adjustment is particular to the sample targeted. So far only a flat
prior has been set for the allowed redshifts; this is given by the search
ranges. However, there are far fewer galaxies at $z>0.5$ than at lower
redshifts in the GAMA survey.  This adjustment is given by
\begin{equation}
{\rm adjust_2} = 
\begin{array}{l} - 0.8 \\ - 4.0 (z - 0.45) \\ ~~~~0 \end{array}
\mbox{~~~for~~~}
\begin{array}{l} z > 0.65 \\ 0.45 < z < 0.65 \\ z < 0.45 \end{array}
\mbox{~~~,}
\end{equation}
and thus
\begin{equation}
\ccfom = {\rm cc_{fom,prelim}} + {\rm adjust_1} + {\rm adjust_2}
\mbox{~~~.}
\end{equation}


Finally, contamination by solar-system light was checked on a tile by tile
basis,\footnote{A tile refers to the set of spectra from 
a single AAOmega configuration.} 
by looking for an excessive number of G-star template matches.  The
number of stars observed as part of our main survey is about two per cent
\citep{baldry10}.  When there was a clustering of ten or more matches to
templates 7--10 (best or second-best redshift estimate), the tile was checked
for solar-system contamination. This was caused by moonlight, under conditions
where scattering had significant structure on the sub-tile scale, with one
notable exception. A cluster of about twenty matches was found to be centred
on the location of Saturn at the time of the observation.  The spectrum of
Saturn can be seen clearly in a few fibres; \textsc{autoz} does not have this
template but picks up the reflected solar absorption lines. In total, 20 tiles
were flagged as having possible solar contamination, and any best redshift
match to templates 7--10 was given a $\ccfom$ value of 2.5. This was applied
to about 300 spectra and these were not included in the redshift confidence
estimation.  In most cases, the targets were re-observed.


The calibration of $\ccfom$ to a redshift confidence was made by comparing
\textsc{autoz} redshifts between different spectra of the same target position
(matched spectra).\footnote{Observations of the same target taken with
  different 2dF configurations are of the same position on the sky within the
  accuracy of the fibre placement, which is $0.3''$.  This is significantly
  less than the fibre diameter, which subtends $2''$ on the sky.}  Redshift
measurements were considered to be in agreement if they were within 450\,km/s
[$\Delta \ln(1+z) < 0.0015$] and in disagreement otherwise. The probability of
agreement is taken to be
\begin{equation}
p_{\rm agree} = p(\ccfomi) p(\ccfomj)
\end{equation}
where the function $p()$ gives the probability each spectrum is correct as
function of the FOM.  This ignores the small chance that both redshifts are
incorrect but in agreement with each other or that both redshifts are correct
but are of different superimposed sources.  The second effect was noticeable
prior to suppression of the solar-system contamination.

\begin{figure}
\centerline{
\includegraphics[width=\singlecolsize\textwidth]{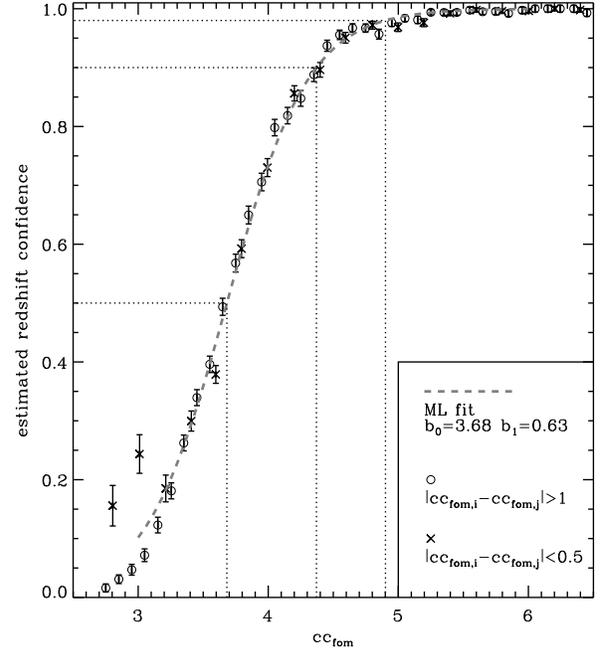} 
}
\caption{Estimate of the redshift confidence as a function of $\ccfom$. The
  circles represent the first binned sample where the FOM values
  between the matched spectra differ by more than 1.0, while the crosses
  represent the second binned sample where the FOM values are similar.  The
  dashed line shows the parametric fit using a maximum likelihood method on
  the unbinned data. The dotted lines show the FOM values corresponding to
  redshift confidence of 0.5, 0.9 and 0.98.}
\label{fig:fom-calibration}
\end{figure}

In order to estimate $p(\ccfom)$, two binned samples were considered.  The
first sample uses the matched spectra where one of the FOM values is higher by
\new{more} than 1.0, and the sample is binned by the lower FOM value. In each bin
the estimate of $p(\ccfom)$ is then $N_{\rm agree}/N$, which assumes the
redshift is correct for the spectrum with the higher FOM value. The second
sample uses the matched spectra where the FOM values are within 0.5 of each
other, and the sample is binned by the mean value. In each bin the estimate of
$p(\ccfom)$ is then $\sqrt{N_{\rm agree}/N}$, which assumes $p(\ccfom)$ is the
same for both spectra in a matched pair. Fig.~\ref{fig:fom-calibration}
shows these binned estimates for the function.

The data were also fitted using a tanh function
\begin{equation}
  p(x) = 0.5 \tanh \left( \frac{x - b_0}{b_1} \right) + 0.5\mbox{~~.}
\label{eqn:tanh}
\end{equation} 
The best fit parameters were determined by maximising the likelihood:
\begin{equation}
\begin{array}{rcl}
 \ln P & = & \Sigma_{i,j,{\rm agree}} \ln [p(\ccfomi) p(\ccfomj)] \: + \\
   & & \Sigma_{i,j,{\rm disagree}} \ln [1 - p(\ccfomi) p(\ccfomj)]
\end{array}
\label{eqn:ml}
\end{equation}
where the summations are over matched spectra with redshift agreement
and disagreement. The best fit was found using only the matched
spectra where the lower FOM value within each pair was between 3.0 and
6.0. This is shown by the dashed line in Fig.~\ref{fig:fom-calibration}. 

Various subsamples of the matched spectra were also considered including a
randomly selected repeat sample (these were targets that were chosen for
reobservation regardless of their assigned redshift quality), and subsamples
where one of the spectra had a particular best-match template. The calibration
points for the subsamples, at which the redshift confidence was 0.9, varied
from $\ccfom \simeq 4.2$ to 4.6. The notable exception was where one of the
matched spectra had a stellar template match. Here there were a higher
fraction of disagreements at $\ccfom>5$ than in other subsamples (0.9
confidence at 5.2). This is a result of star-galaxy blends and is not of
concern for the \textsc{autoz} method. The GAMA target selection is for
extended sources, and thus it is not surprising that a higher fraction of
targets with stellar redshifts are part of a star-galaxy blend compared to
random selection; in addition spectra of star-galaxy blends are more likely to
have been reobserved given the increased difficulty of assigning redshifts
using \textsc{runz}. Overall, we use $b_0=3.7$ and $b_1=0.7$ to assign
redshift confidence, which is slightly more conservative than the best fit to
all the matched spectra (Fig.~\ref{fig:fom-calibration}).

\begin{figure}
\centerline{
\includegraphics[width=\singlecolsize\textwidth]{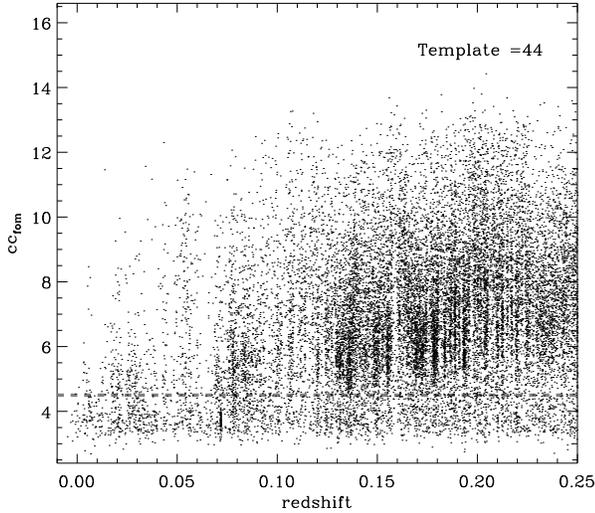} 
}
\caption{Example diagnostic plot of $\ccfom$ versus redshift.
  The dashed line shows the standard cut to select sufficient quality
  redshifts.}
\label{fig:diagnostic-z-ccfom}
\end{figure}

After the redshift confidence calibration, 248\,145 spectra were assigned a
probability $p(\ccfom) > 0.9$ (AATSpecAutozAllv22, which uses the galaxy
templates 40--47). The mean calibrated confidence using this quality selection
is 0.996 implying that less than 1\% would be assigned an incorrect
redshift.  For this and earlier versions, additional diagnostics were run to
check for anomalies including: checks for solar-system contamination
(described earlier), plots of spec-$z$ versus photo-$z$ for every tile, plots
of matched spectra where there was redshift disagreement with $\ccfom>5$ for
both, and plots of $\ccfom$ versus $z$ for each template.  An example of the
latter is shown in Fig.~\ref{fig:diagnostic-z-ccfom}. The cut at $\ccfom=4.5$
corresponds approximately to our standard quality cut ($p>0.9$). One can see
the redshift spikes corresponding to large-scale structure above the line,
with more scatter below the line. There is a narrow artefact below the line at
$z\simeq0.0715$. In previous iterations of the \textsc{autoz} code, poor sky subtraction, for
example, was evident with signatures of serious artefacts extending from below
to above the quality cut. These artefacts were eliminated by changes to the
code (e.g., broadening of the error spectrum), and by improved reduction of
several tiles.

\subsection{Redshift completeness of 
  the GAMA equatorial fields}

The large number of targets with two or more spectra taken has allowed the
\textsc{autoz} code to be accurately calibrated.  However, the main aim of the
repeated observations was simply to obtain high redshift completeness. If a
target was first observed in poor conditions or using a fibre with lower than
average transmission efficiency, then the second observation was often
successful in obtaining a redshift.

\begin{figure}
\centerline{
\includegraphics[width=\singlecolsize\textwidth]{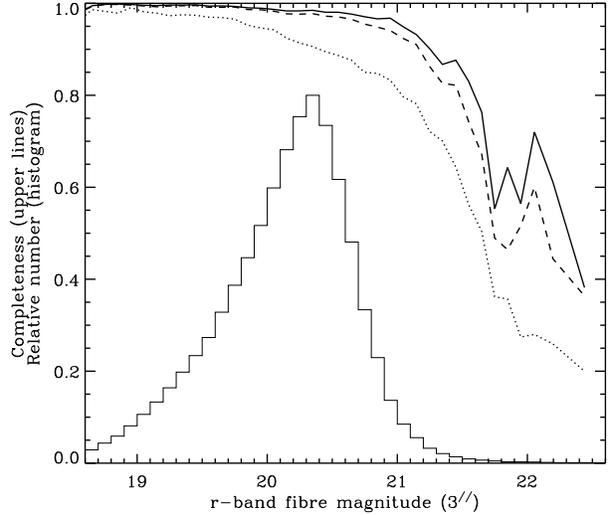} 
}
\caption{Redshift completeness as a function of fibre magnitude for
  main survey targets (top three lines) using: the first observation
  of any target (dotted line), the best observation (dashed line), and
  incorporating coadded observations (solid line). The histogram of
  fibre magnitudes is also shown (lower solid line).}
\label{fig:comp-progress}
\end{figure}

To demonstrate the effect of the strategy on the redshift completeness, we
select the sample of main sample targets in the G09, G12 and G15 that have
been observed at least once with AAOmega (as per
Fig.~\ref{fig:fibre-histo}).\footnote{The main sample targets are
  predominantly $r_{\rm petro}<19.8$ with good visual classification.  The
  sample has been cleaned of SDSS database objects where the photometry looks
  to be significantly in error. The number of main sample targets is 191\,051
  in the equatorial fields, of which, 173\,164 have AAOmega
  observations (TilingCatv42).}
Fig.~\ref{fig:comp-progress} shows the redshift completeness as a function
of fibre magnitude.  The dotted line shows the completeness ($p>0.9$) if only
the first observation is used for each target (or a pre-existing good
redshift), while the dashed line shows the completeness using the best
observation. Note the significant improvement when poor quality observations
are replaced.

For some targets that are intrinsically faint, a further improvement can be
made by coadding any repeated observations.  In principle, spectra should be
coadded in proportion to their signal divided by their noise squared. In
reality, it is hard to estimate the signal because of the faintness and
sometimes poor sky subtraction. Since we will only obtain a significant
improvement in the FOM if the signal-to-noise ratio 
is similar for two spectra that
are summed together, we assume that this is the case.  The spectra to be
summed were high-pass filtered and normalised, as per
Fig.~\ref{fig:example-hpf} and \S~\ref{sec:aaomega-spectra}, separately. The
HPF spectra were then rebinned onto a heliocentric vacuum wavelength scale
before summing. The \textsc{autoz} cross-correlation and redshift confidence
estimation was then run on the summed HPF spectra. The solid line in
Fig.~\ref{fig:comp-progress} shows the resulting increase in completeness when
low confidence single-spectrum redshifts are replaced by coadded-spectrum
redshifts where there is an improvement in the FOM. Overall, the fraction of
the GAMA main sample with $p>0.9$ \textsc{autoz} redshifts is 98.3\%.  The
mean $p$-value after selecting on this redshift quality cut is 0.992.

\section{Redshift uncertainties}
\label{sec:uncertainty}

In this section, we discuss the measurement uncertainty 
assuming that the correct 
redshift peak has been assigned. This uncertainty comes from errors in
wavelength calibration, noise affecting the centroid of the peak and mismatch
between the template and observed spectra.  To test this we take the sample of
matched spectra where the redshifts are in agreement (within 450\,km/s) and
both have $\ccfom>3.5$ but with the redshift peaks coming from {\it different}
galaxy templates. This gives a sample of 12\,323 matched spectra.

Following \citet{TD79} and from trial and error, a reasonable predictor of the
ability to centroid is given by 
\begin{equation} {\cal V} = \frac{v_{\rm fwhm}}{1 + r_x} \end{equation} 
where $v_{\rm fwhm}$ is the velocity full-width half maximum of the peak.  The
latter is determined from the number of rebinned pixels
(13.8\,km/s) within 600\,km/s of the peak above the half maximum of the
cross-correlation function. This is more robust and less time consuming than
fitting a Gaussian or other function to the peak. Overall, we model the
variance in the velocity (redshift) uncertainty as:
\begin{equation}
  \sigma_v^2 = \left(c \frac{\sigma_{z}}{1+z} \right)^2 = 
  c_0^2 + c_1^2 {\cal V}^2 \mbox{~~~.}
\label{eqn:sigma-v}
\end{equation}
Thus the expected mean square of the velocity difference between matched
spectra $i$ and $j$ is given by
\begin{equation}
   \langle \Delta v_{i,j}^2 \rangle = 
   2 c_0^2 + c_1^2 \left( {\cal V}_i^2 +
     {\cal V}_j^2 \right)
\mbox{~~~.}
\label{eqn:velocity-fit}
\end{equation}

\begin{figure}
\centerline{
\includegraphics[width=\singlecolsize\textwidth]{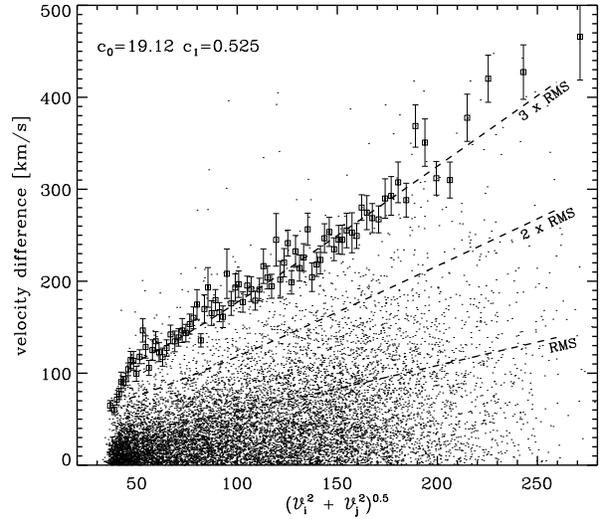} 
}
\caption{Velocity difference versus $\sqrt{{\cal V}_i^2 + {\cal V}_j^2}$. The
  points represent the $|\Delta v_{i,j}|$ values for the matched spectra.  The
  squares, with error bars, show {\it three times} the RMS velocity
  differences in bins.  The dashed lines show one, two and three times the RMS
  values using a fit for $c_0$ and $c_1$ (Eq.~\ref{eqn:velocity-fit}) between
  45 and 260.}
\label{fig:velocity-errors}
\end{figure}

\begin{figure}
\centerline{
\includegraphics[width=\singlecolsize\textwidth]{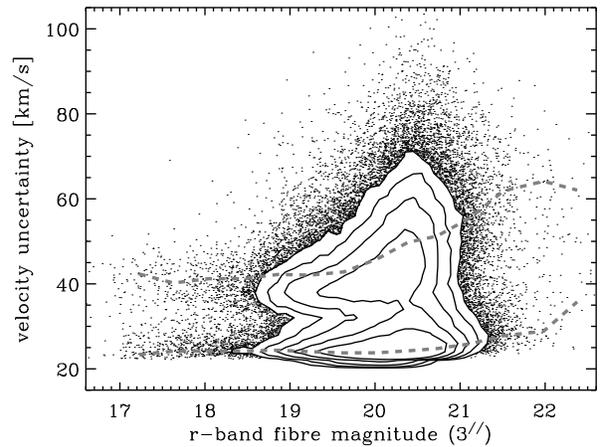} 
}
\caption{Calibrated velocity uncertainty as a function of the fibre
  magnitude. The GAMA-AAOmega sample distribution is shown by the
  points and solid-line contours, for the equatorial fields, 
  while the 16th and 84th percentiles
  in bins of the fibre magnitude are shown by the dashed lines.
  The lower ridge corresponds primarily to spectra matched to templates with 
  strong emission lines, while the ridge at about 40\,km/s corresponds 
  to spectra matched to templates with weak or no emission lines.}
\label{fig:fibre-velocity}
\end{figure}

Fig.~\ref{fig:velocity-errors} shows the velocity difference ($|\Delta
v_{i,j}|$) versus the predictor value $\sqrt{{\cal V}_i^2 + {\cal V}_j^2}$.
This shows that the mean velocity difference increases with average ${\cal V}$
as expected.  The mean square values of the velocity differences were
determined in bins of the predictor value.  These binned values are shown in
the figure (plotted as three times the RMS value) and were used to determine
$c_0$ and $c_1$ using a fit between 45 and 260.  Over this range, the fit is
good and thus is expected to give a reasonable estimate of $\sigma_v$. The fit
overpredicts the velocity difference at predictor values less than 45. These
correspond to spectra with strong emission lines. For the purposes of group
dynamical measurements, it is more important to have an accurate measure of
the velocity uncertainty when this uncertainty is larger and thus we apply the
fit to all the GAMA-AAOmega spectra. For the high quality redshift sample
($p>0.9$), the median $\sigma_v$ is 33\,km/s, with
85 per cent of redshifts having an uncertainty less than 50\,km/s using this
calibration. Fig.~\ref{fig:fibre-velocity} shows the distribution 
of velocity uncertainty as a function of the fibre magnitude. 

\subsection{Comparison with SDSS DR10}

At the start of the GAMA spectroscopic campaign, SDSS DR7 and other
pre-existing redshifts were matched to the input catalogue
\citep{baldry10}. Since then SDSS has had three more data releases, including
spectra from the BOSS survey \citep{dawson13}. In addition, the primary method
of determining redshifts was updated to $\chi^2$ fitting using eigenspectra
\citep{bolton12}. This was applied retrospectively to all plates observed for
the SDSS surveys.
 
In order to compare GAMA \textsc{autoz} redshifts with SDSS, we selected all
SDSS DR10 redshifts in the GAMA fields with \textsc{zwarning}=0 (primarily a
reduced $\chi^2$ difference of \new{more} than 0.010 between the 1st and 2nd
redshift peaks).  This corresponds approximately to $p>0.9$ from our tests,
and as implied by \citet{bolton12}'s test, which showed that as the $\chi^2$
difference threshold was lowered to 0.008, 8\% of the additional redshifts
were estimated to be incorrect.  Matching the SDSS \textsc{zwarning}=0 spectra
from DR10 to the GAMA AAOmega spectra ($p>0.9$, excluding spectroscopic
standards) within 1\,arcsec results in 9426 cross matches.  From these, 99.1\%
have redshifts in agreement between GAMA and SDSS, which is as expected for a
mean confidence of 99.5\% for the GAMA sample and a similar value for the SDSS
sample.

\begin{figure}
\centerline{
\includegraphics[width=\singlecolsize\textwidth]{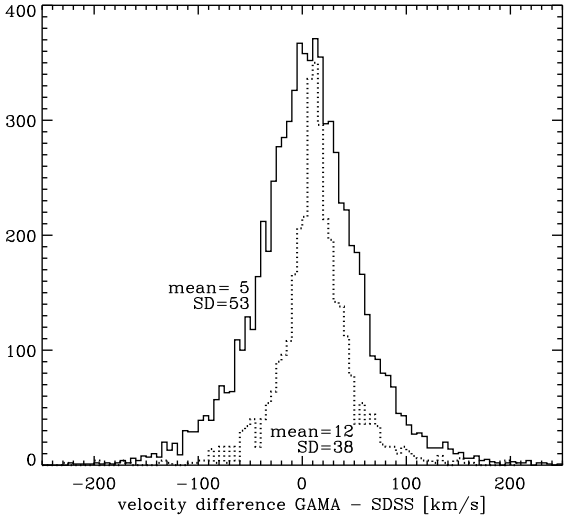} 
}
\caption{Velocity difference between GAMA and SDSS redshift
  measurements. The solid line shows the histogram for the SDSS-BOSS sample
  of galaxies, while the dotted line shows the histogram for the
  SDSS-legacy sample (histogram values scaled up by a factor of two).}
\label{fig:gama-sdss-veldiff}
\end{figure}

There are 1748 matches to spectra from the original SDSS spectrographs that
were part of the legacy survey (main galaxy sample and luminous red galaxies).
These matches were obtained for quality control purposes or because SDSS had
reported a low redshift confidence in DR7.  There are 7674 matches to spectra
from the BOSS spectrographs and survey.  The large number of matches is a
result of GAMA and BOSS independently choosing these targets.
Fig.~\ref{fig:gama-sdss-veldiff} shows the velocity difference histogram for
the two samples [$\Delta v_{i,j} = c \ln(1+z_{\rm GAMA}) - c \ln(1+z_{\rm
  SDSS})$].  For the legacy sample, the mean and standard deviation were 12
and 38\,km/s, and for the BOSS sample, they were 5 and 53\,km/s. The standard
deviations are as predicted from the velocity errors estimated above for GAMA,
with a smaller contribution from SDSS as estimated by \citet{bolton12}.  The
larger mean offset with respect to the legacy sample may be because of the
different eigenspectra used: spEigenGal-53724.fits as opposed to
spEigenGal-55740.fits used by BOSS and GAMA \textsc{autoz}.  Other than this
small anomaly, the comparison with SDSS supports the GAMA estimates of
redshift confidence and velocity uncertainty described in this paper.

\section{Summary}
\label{sec:summary}

We have developed a redshift measurement code called \textsc{autoz} for use on
the GAMA AAOmega spectra. The method uses the cross-correlation technique with
robust high-pass filtering suitable for galaxy and stellar types applied
to the templates (Fig.~\ref{fig:hp-procedure}) and observed spectra
(Fig.~\ref{fig:example-hpf}). The observed HPF spectra are inversely weighted
by the variance estimated at each pixel, broadened slightly by a maximum
kernel filter.  To avoid giving too much weight to emission line matches,
large deviations in the HPF spectra are partially clipped for both the
observed spectra and templates (Fig.~\ref{fig:template26}). Lowering the
weight of large deviations reduces the impact of spurious peaks caused by
uncorrected artefacts.  Real cross-correlation peaks are rarely adversely
affected by this because there is additional signal at the correct redshift
peaks from other parts of the spectra.

For each observed spectrum, the cross-correlation functions are determined for
every chosen template. Each function is normalized by dividing by the RMS of
the turning points over a specified noise-estimate range
(Table~\ref{tab:templates}).  These ranges were chosen so that the value of a
peak represents a similar confidence level across all the templates.  The best
four redshift estimates are obtained from the cross-correlation function peaks
(Fig.~\ref{fig:cross-correlation}), not including peaks within 600\,km/s of a
better redshift estimate.  A FOM for the redshift confidence is determined
using the value of the highest peak ($r_x$), and the ratio of $r_x$
to the RMS of the 2nd, 3rd and 4th peaks
(Eqs.~\ref{eqn:ccsigma1to234}--\ref{eqn:fom-prelim},
Fig.~\ref{fig:fom-calc}). Overall, the procedure can be adjusted and the FOM
calibrated using repeat observations within a survey.

The GAMA AAOmega redshift survey has taken spectra of over 230\,000 unique
targets. As part of a strategy of obtaining high completeness and for quality
control, about 40\,000 targets have had two or more spectra taken. These
repeats were used to calibrate the confidence level as a function of the FOM
using a maximum likelihood method (Eqs.~\ref{eqn:tanh}--\ref{eqn:ml},
Fig.~\ref{fig:fom-calibration}).  Overall, the \textsc{autoz} code has
significantly improved the redshift reliability within the GAMA main sample,
with a high completeness for targets with $3''$-aperture $r$-band magnitudes
as faint as 21\,mag (Fig.~\ref{fig:comp-progress}).  The redshift uncertainties
have also been calibrated using the repeat observations, with most having
redshift errors less than 50\,km/s (Fig.~\ref{fig:fibre-velocity}).
\new{\textsc{autoz} measurements will be included in public data releases 
  from GAMA DR3 onwards.}

With some consideration to making adjustments --- templates, high-pass
filtering scale, clipping limits, noise-estimate ranges, FOM calculation and 
calibration --- the fully automatic method outlined here could be used 
for other large galaxy redshift surveys. A key factor is using a
sufficient number of repeats, both random and at the fainter end of the sample, 
to allow for an accurate empirical confidence calibration. 

\section*{Acknowledgements}

Thanks to the anonymous referee for helpful comments on the manuscript. 
The \textsc{autoz} code makes use of routines from the 
IDL Astronomy User's Library. 
GAMA is a joint European-Australasian project based around a spectroscopic
campaign using the Anglo-Australian Telescope. The GAMA input catalogue is
based on data taken from the Sloan Digital Sky Survey and the UKIRT Infrared
Deep Sky Survey. Complementary imaging of the GAMA regions is being obtained
by a number of independent survey programs including GALEX MIS, VST KiDS,
VISTA VIKING, WISE, Herschel-ATLAS, GMRT and ASKAP providing UV to radio
coverage. GAMA is funded by the STFC (UK), the ARC (Australia), the AAO, and
the participating institutions. The GAMA website is
http://www.gama-survey.org/ .

\setlength{\bibhang}{2.0em}
\setlength\labelwidth{0.0em}




\label{lastpage}

\end{document}